\documentclass{aa}
\usepackage{epsfig,times}
\usepackage{amssymb}
\usepackage{txfonts}
\usepackage{graphicx}
\usepackage{natbib}
\begin{document}

\title{Confirmation of the occurrence of the Hall instability in the non-linear regime.  }

\author{J.A.~Pons\inst{1}  \and U.~Geppert\inst{2}}
\institute{Departament de F\'{\i}sica Aplicada, Universitat d'Alacant,  
Ap. Correus 99, 03080 Alacant, Spain
\and 
German Aerospace Center, Institute for Space Systems, Rutherfordstr. 2, 12489 Berlin, Germany}
\date{Received...../ Accepted.....} 
 
\abstract 
{The non-linear Hall term present in the induction equation in the electron-magneto-hydrodynamics limit 
is responsible for the Hall drift of the magnetic field and, in some cases, for the so-called Hall instability. }
{We investigate whether or not the growth rates and eigenfunctions found in the linear analysis
are consistent with the results of non-linear numerical simulations.}
{Following the linear analysis of Rheinhardt \& Geppert, we study the same cases for which the 
Hall instability was predicted by solving the non-linear Hall induction equation using a two-dimensional 
conservative and divergence-free finite difference scheme that overcomes intrinsic difficulties of 
pseudo-spectral methods and can describe situations with arbitrarily high magnetic Reynolds numbers. }
{We show that unstable modes can grow to the level of the background field without being
overwhelmed by the Hall cascade, and cause a complete rearrangement of the field geometry. 
We confirm both the growth rates and eigenfunctions found in the linearized analysis and hence the instability. 
In the non-linear regime, after the unstable modes grow to the background level, the naturally selected 
modes become stable and oscillatory. Later on, the evolution tends to select the modes with the longest 
possible wavelengths, but this process occurs on the magnetic diffusion timescale. }
{We confirm the existence of the Hall instability. We argue against using
the misleading terminology  that associates the non-linear Hall term with a turbulent Hall cascade, since small-scale 
structures are not created everywhere. The field evolves instead in a Burgers-like manner, forms local 
structures with strong gradients which become shocks in the zero resistivity limit, and Hall waves are launched and
propagated through the entire domain.}

\titlerunning{Confirmation of the Hall instability in the non-linear regime} 
\authorrunning{J.A. Pons \& U. Geppert} 
 
\maketitle

\section{Introduction}
The influence of the Hall term  on the magnetic field evolution of a neutron star (NS) has been debated 
for many years. By analogy with the vorticity equation in ordinary hydrodynamics, \cite{GR1992} conjectured that the transfer 
of magnetic energy from large to small scales proceeds in a similar way to ordinary turbulence. 
However, the analogy of the Hall induction equation with the vorticity equation
is not complete, and the conjecture remained to be confirmed by multidimensional numerical simulations.
\cite{Vai2000} proposed a mechanism for the fast dissipation of magnetic field based on the Hall drift in 
stratified media. They correctly pointed out that Hall currents are able to create current sheets (which are sites
for efficient dissipation) and that the evolution of the toroidal field resembles the Burgers equation. The
same Burgers-like equation is applicable even to non-stratified media, but in a spherical shell \citep{PonsGeppert2007}, in which 
the Hall term in the induction equation  tends to create current sheets instead of ordinary turbulence.
\cite{RG2002} (hencefort RG) showed by a linear analysis that, in a one-component (electron) plasma, a 
large-scale background magnetic field may become unstable to smaller scale perturbations.
This Hall-drift induced instability (HI) occurs when the
magnetization parameter is high and the background field has enough curvature. Since these conditions may 
be realized in the crust of a NS, the problem of the HI became interesting not only from the MHD point of view but also for 
the astrophysics community. 
Although the Hall-drift is a non-dissipative process, the growth of small-scale magnetic field 
components modifies the overall magnetic field structure 
and opens the possibility of more rapid field decay than pure ohmic dissipation would predict. 
This effect may be relevant, at least, in relatively
young neutron stars.  Determining its effect on the long-term evolution, and which configurations are stable on long time-scales
are different issues. A review by \cite{Reis2009}
discusses the main properties of stable magnetic configurations emphasizing the importance of a non-barotropic stratification
in neutron stars, and provides insight into the expected long-lived magnetic field structures.
Some observational implications of rapid magnetic field decay were also discussed by
\cite{GR2002,PonsLink2007,Aguilera2008a,Aguilera2008b}. Other attempts
to understand the long-term evolution of neutron star magnetic fields by numerical simulations
were performed by \cite{Hoyos2008}, but were restricted to 1-dimensional geometries.  

The appearance of the HI was disputed by \cite{HR2002}: they did not find any evidence of the instability 
in their numerical study of field decay under the influence of the Hall drift in a constant density 
spherical shell. The study was later extended to include density gradients
\citep{HR2004}. However, it was clear that they could not find the HI because of numerical limitations
of their code, which is unable to handle situations with sufficiently high magnetic Reynolds number.
The relevance of the HI to astrophysical scenarios had already been questioned 
by RG. Later \cite{CAZ2004} conjectured that the gap in the energy spectrum created by the HI could be rapidly filled in 
by the Hall cascade. They also computed Hall-wave eigenfunctions, including the effect of density gradients across the crust,
and found agreement with the results of RG. An important component of their study was their estimate of the response of the crust 
to the magnetic stresses induced by Hall waves.
\cite{PonsGeppert2007} performed 2D simulations by solving the non-linear Hall-induction equation
in a realistic NS crust and found some evidence for a rapid reorganization of the field 
during a relatively short period, but no clear confirmation of the HI could be concluded. The code again had 
numerical limitations that avoided additional exploration of extreme limits to confirm or rule out the existence
of the HI. 
\cite{WH2009}(hereafter WH) used their 3D pseudo-spectral 
code and concluded that the HI proposed by RG is  a consequence of particular initial conditions and 
irrelevant compared to the Hall cascade. They considered a similar, but not identical,
geometry as RG and used a code limited to employing periodic boundary conditions in all directions.
 
In this Letter, we solve the non-linear induction equation using a new code 
{(Pons et al., 2010, in preparation)} based on a divergence-free, conservative, finite-difference 
scheme that resembles numerical methods widely employed in astrophysical MHD problems to handle situations where 
shocks are present. 
It solves problems caused by the transition from a parabolic equation to the hyperbolic regime as the 
magnetic resistivity vanishes. 
Since the code is designed by following methods developed to solve hyperbolic equations, which are similar to standard
techniques used in perfect fluid hydrodynamics and ideal MHD \citep{Anton2006}, it  is able to deal with zero physical resistivity,
therefore extending all previous 
studies to the limit of arbitrary high magnetic Reynolds numbers. Our numerical scheme leads to a fourth order numerical
hyper-resistivity, in a similar way to other studies on the basis of which we developed our own algorithm \citep{OD2006,Toth2008}.
We performed a number of tests including the purely diffusive case and the exactly non-resistive case (e.g., propagation of whistler waves on a
homogeneous background). In all cases  where analytical solutions are available,
the correct modes, decay times, and wave propagation speeds were reproduced.

This allows us to study the same model originally analyzed by RG and answer open questions about the 
occurrence of the HI and its interplay with the so-called  {\it Hall cascade}. 

\section{Basic equations and model set up.}
To be able to compare with the results of  RG, we restrict ourselves to the case of constant density 
and electric conductivity. In dimensionless variables (for details, see RG),
the Hall induction equation is given by
\begin{equation}
\frac{\partial\vec B}{\partial t}=\Delta\vec{B} - \nabla\times\left[(\nabla\times\vec{B})\times\vec{B}\right]\;,
\label{Hallind_norm}
\end{equation}
which must be solved  with the constraint $\nabla \cdot \vec{B}=0$.

Following RG, we consider a slab that is infinitely extended in both $x-$ and $y-$directions. 
The upper boundary is located at $z=+1$ in contact to vacuum, while the lower boundary, located at $z=-1$, 
forms the interface with a perfect conductor. 
The background (initial) magnetic field is assumed to have only the $x-$component and depends only on the $z-$coordinate 
$\vec{B}_0 = B_0 f(z) \vec{e}_x$, where $\vec{e}_x$ is the unit vector in the $x-$direction,
$B_0$ is a constant fixing the maximum field strength, and $f(z)$ is an arbitrary function. 
The reason for this choice is that we must start with a background field that is not affected by the non-linear Hall term. 
Note that this is not an arbitrary choice: in any problem where 
one wants to study the  evolution of perturbations on a background, the first important requirement is that 
the background has to be an equilibrium solution. Any other choice of
background field, will cause rapid dynamical evolution.
After all, nobody would attempt, for instance, to study perturbations within a star starting with a fluid configuration in which gravity 
is not balanced by the gradient of pressure. Thus, imposing the condition  ${\nabla \times}[(\nabla \times \vec{B}_0) \times \vec{B}_0]=0$,
which was criticized by WH, is a pre-requisite for a linear analysis, not just a special choice. 

A different question is whether or not this background is an accurate choice and how this affects the astrophysical relevance of the conclusions.
Naively, since a NS at birth is fluid, one would expect that an MHD equilibrium is reached before the crust is formed, and thus that the initial configuration
may be closer to an MHD equilibrium than to a Hall equilibrium. But the conditions for reaching these two equilibria are similar (not identical) and
one may expect that the initial model is not far from a Hall equilibrium.
Another astrophysically interesting scenario could be relatively old NS, with an age of about $10^6$ years. 
During the photon cooling era, the temperature can become so low that the magnetization parameter becomes high even for moderate
magnetic fields ($10^{12}-10^{13}$ G), but it does take time for the temperature to reach such low values. 
The system slowly approaches slowly the non-resistive case,  for a relatively long period not being affected by the HI. 
It is reasonable to assume that the crustal magnetic field has enough time to
steadily adjust itself to a Hall equilibrium until the
conditions for triggering the instability can be reached in a similar manner as described above.

The choice of the function $f(z)$ is of course not unique. To be able to compare with RG, we use 
\begin{equation}
f(z)=(1+z)(1-z^2),
\label{fz}
\end{equation}
for which the modes with
the most rapid growth rate correspond to horizontal wavenumbers $k_y=0$ and $k_x$ of order 1 for $B_0\approx 1000$. 
Therefore, we restrict ourselves to the 2D case with $k_y=0$, e.g., we assume independence on
the $y-$coordinate. 

All quantities are assumed to depend periodically on $x$ with period $L$
(that is, the length of the numerical domain in the x-direction), and periodic boundary conditions are 
applied in terms of the $x-$coordinate.
For the inner boundary condition, as in former studies, we mimic the conditions in a NS's crust by 
assuming that the matter beneath the lower surface at $z=-1$ is a perfect conductor. In that case, the 
Meissner-Ochsenfeld effect prevents the magnetic field from penetrating into the superconducting matter. 
This requires that the normal component of the magnetic field and the tangential components of the electric field 
have to vanish \citep{PonsGeppert2007}, i.e., we assume that $B_z=E_x=E_y=0$ at that interface.
At the transition to a vacuum ($z=+1$), the analog to the toroidal component $B_y$ has to vanish, but 
the boundary conditions for the other components are not so simple.
In terms of individual modes with fixed wavenumber, a spectral representation that matches the vacuum solution 
needs to satisfy a simple relation between the Stokes function and its 
derivative (RG). However, in real space and using the components of the magnetic field, 
we have to employ a non-local boundary condition. 
By using Green's representation formula to solve the Laplace equation in a bounded domain, and
considering that we reduce the problem to 2D, one can derive an explicit form for the outer boundary condition 
(Rheinhardt 2009, private communication) given by
\begin{eqnarray}
B_x(x)
&\approx& \frac{1}{\pi}\int_0^L{B_z(\xi)}\left(\sum_{k=-N}^{N}\frac{1}{x-(\xi+kL)}\right)d\xi\;\;, ~{\rm at}\; z=+1
\end{eqnarray}
where the integer $N>0$ must be chosen to be high enough to ensure that the sum converges. 

With the same initial model and physically motivated boundary conditions, RG found that the HI appears for $B_0 \gtrsim 3$, and 
that some small-scale structures must grow quickly. Since there is some confusion in the literature about the 
interpretation of the term {\it small scales}, we briefly review their results and make some remarks.
For sufficiently strong and curved background fields, there is always a region in the wavenumber space where unstable 
modes appear. The highest growth rates correspond to $k_y=0$ and 
the range of $k_x$ allowing for unstable modes changes depending on both the form of $f(z)$ and
the value of $B_0$ but,  in dimensionless units and with 
the slab defined in the region $z=[-1,+1]$, it is in all cases of order 1, or even smaller.
This means that {\it the most rapidly growing unstable mode has a typical 
wavelength of the order of the slab height}. Thus, in some sense, it cannot be rigorously assumed to be 
{\it small-scale mode}. However, we note that the typical crust size of a NS is about 1 km, compared to its radius 
of about 10 km. Thus, if the HI occurs, one would expect structures of a typical size of one km to grow within 
the crust, with a typical area of $\approx 10^3$ km$^2$. In this respect, the field geometry will 
differ considerably from a typical dipolar field, consisting instead of tens or hundreds of smaller structures, 
and that is what RG meant by small-scale structures. In addition, the toroidal field (here $B_y$) may grow rapidly in a 
very narrow region of 10-100 m beneath the surface. This is far 
from the typical picture of turbulence in fluid dynamics, where multiple small-scale substructures grow in a large region at the 
same time. In the HI, there are a few small-scale structures, which can be very localized, in particular close to the boundaries, 
or wherever the toroidal field changes sign, because the Burgers-like term in the evolution equation will produce a sharp
discontinuity.
This qualitative difference is easy to visualize when one follows the evolution of the field components in the real space.
It becomes, however, more difficult to distinguish when working in wave-number space with pseudo-spectral 
methods if we only consider the power-law in the spectrum. 
We think that most of the disagreement and confusion in the past  few years has been caused by an inaccurate use 
of language, terms such as {\it Hall cascade} or {\it small scale} being interpreted differently  by  different readers.

\section{Results}
\begin{figure*}[th]
\centering
\includegraphics[width=0.9\textwidth]{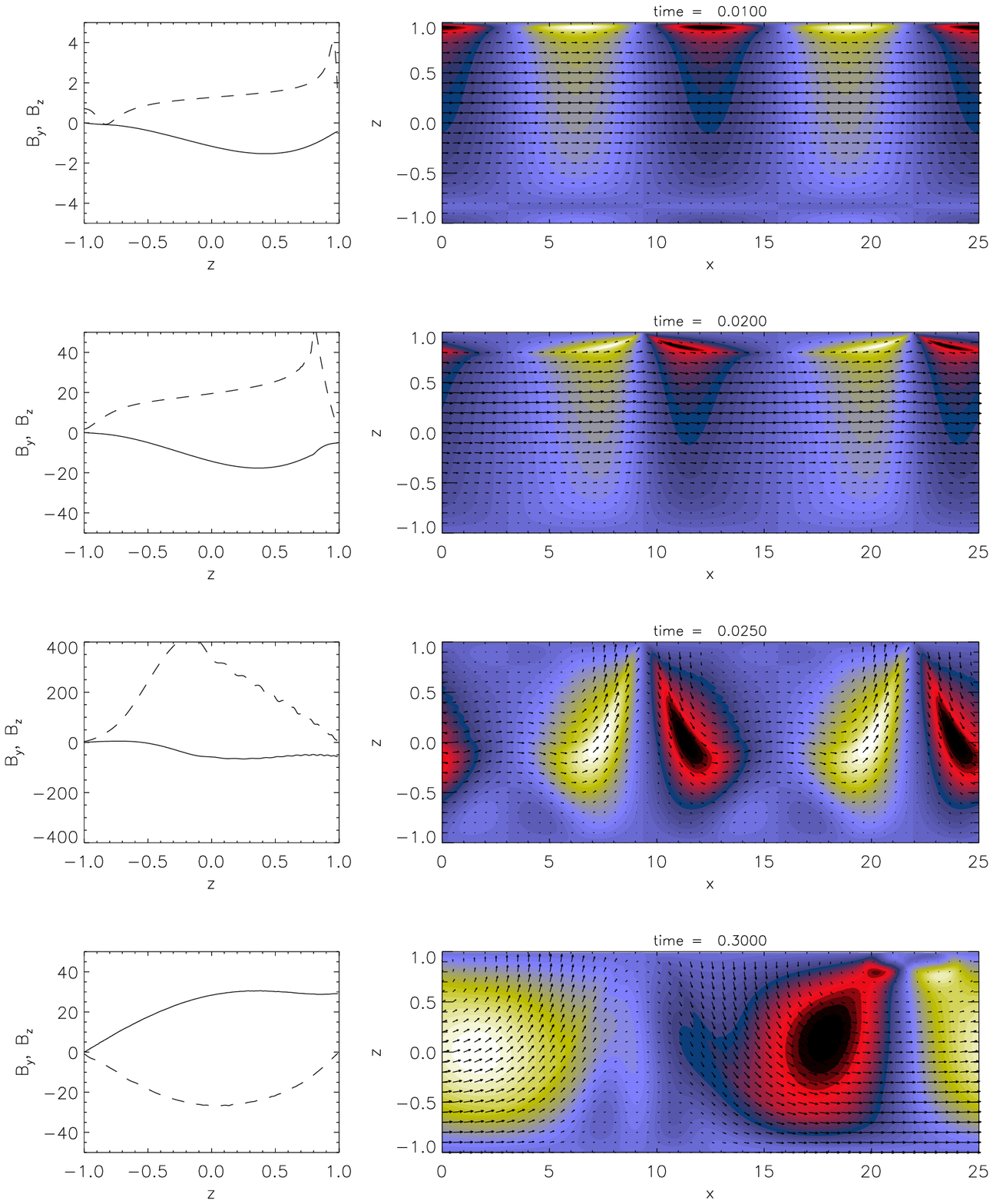}
\caption{Evolution of the initial configuration defined by Eq. \ref{fz}  with 
a $k_x=0.5$ perturbation at four different times.
The left panels show transverse cuts of the $B_y$ (dashes) and $B_z$ (solid line) components at $x=0$, while the right 
panels show the field structure. Arrows show the 
$B_x,B_z$-components and the color scale the $B_y$ component (red/black positive, yellow/white negative).
Note the different scales in the left panels.
}
\label{fig1}
\end{figure*} 
WH claimed that any instabilities are always overwhelmed by a turbulent Hall cascade.
To either confirm or reject these conclusions, we performed simulations following the 
non-linear evolution of the initial model described above. We assumed that $B_0=1000$,
which, according to the linear analysis, should be affected by the HI. The wavenumber of unstable modes is in the range 
$k_x=[0.3,3]$ and growth rates in the range $2\times10^{-3}-10^{-2}$ ohmic decay times. We
perturbed the background field with small perturbations (0.1-1\% in amplitude) of different wavenumbers in 
that range, and followed the time evolution for many growth times, until a quasi-stationary diffusion regime was
reached.
In most cases, we found that the growth of the structures closely resembled the eigenfunctions obtained in
RG with the correct timescales.
Figure \ref{fig1} shows the results for an initial perturbation of $k_x=0.5$ and maximum amplitude 1. 
This mode should be unstable and its growth time is about $2\times10^{-3}$. As predicted,  the initial perturbation 
(which is not a pure  mode eigenfunction) quickly reaches the typical form of the eigenfunctions, and begins
to grow rapidly on the expected timescale.

\begin{figure}[th]
\centering
\includegraphics[width=0.45\textwidth]{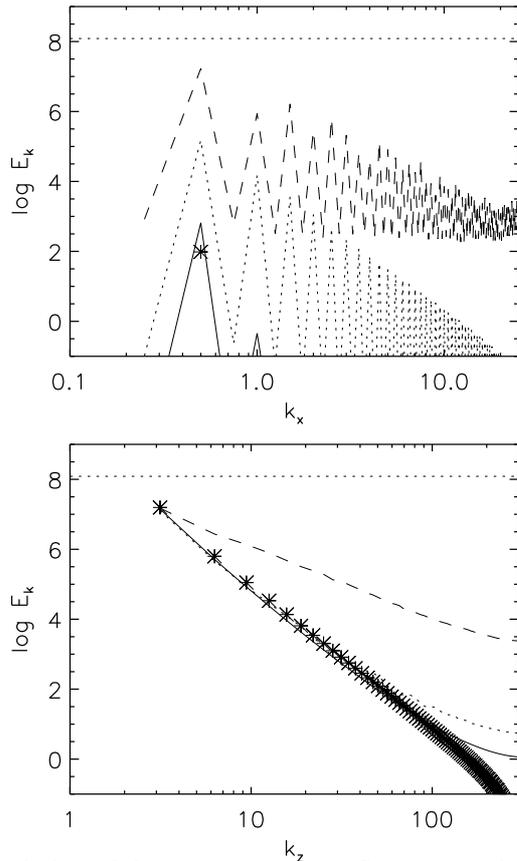}
\caption{Evolution of the power spectrum. Spectra are shown at times $t=0, 0.01, 0.02$ and 0.03 with asterisks, 
solid lines, dots, and dashes, respectively. The total energy is collapsed onto the $k_x$ axis (top) and 
$k_z$ axis (bottom). The upper dotted line 
marks the initial power in the $k_x=0$ background field, from which the energy of the unstable modes is drained.  
}
\label{fig2}
\end{figure}

We note the similarity of the field structure at $t=0.01-0.02$ with Figs. 3 and 4 in RG, which delineate the 
structure of the most rapidly growing mode.  During the linear stage (a few Hall times), 
the unstable modes grow exponentially retaining their structure, as expected, until they reach amplitudes large enough for non-linear
effects to be important. At that point ($ \gtrsim 0.02$), the evolution becomes non-linear, there is a drift towards the inner boundary,
and the global shape of the field is completely reorganized. After a stationary configuration is adopted, we enter
the diffusion regime and observe long-lived, damped oscillatory modes with smaller scale modes that dissipate faster.
During the late evolution (bottom panel of Fig. \ref{fig1}), only the longest wavelength modes survive \footnote{ 
Animations of the long-term simulations are available at
{\tt http://www.ua.es/personal/jose.pons/movies.html}, where the evolution is visualized in much more detail 
than a figure can show.}.
The evolution of the field in terms of the power spectra is shown in Fig. \ref{fig2}. 
The fast growth of the initial perturbation ($k_x=0.5$) and other unstable modes ($k_x=1.0, 1.5, ...$) 
at early times is evident. After the unstable-mode energy has grown several orders of magnitude, non-linear coupling fills the full spectrum 
by $t=0.03$,
but the dominant mode (besides $k_x=0$) is always the $k_x=0.5$ mode, which has drained more than 10\% of the background
energy. Collapsed onto the $k_z$ axis, the spectra only shows the ordinary cascade filling short wavelength modes, but this only indicates 
that the field has drifted vertically, been compressed into a narrower region, and should not be interpreted as
extended turbulence over the whole domain.

In contrast to WH, who never detected the growth of the initial perturbation and thus concluded that it is 
always subsumed into the standard Hall cascade, we observed the same unstable behavior shown in Figs. \ref{fig1} and \ref{fig2}
in a number of other models with different amplitude and $k_x$ of the initial perturbation. The same result is obtained with
different grid resolutions.
We also explored other initial choices for $f(z)$, (e.g. $f(z) \propto \sin(k_z z)$)  and replaced the outer boundary condition with the simpler condition
$B_x=B_y=0$ obtaining qualitatively similar results.
What then is the origin of this discrepancy? 
We note that the finite size ($L$) of the grid fixes the lower 
limit of $k_x$. Periodic boundary conditions result in numerical truncation of all modes with $k_x<2\pi/L$.
In the models in which we restricted the horizontal grid to the interval $[0,2\pi]$ 
(the typical size in WH models), we observed that unstable modes grew initially but the poloidal 
perturbations reached only a few percent of the total energy and the toroidal component saturated even earlier.
Although the linear analysis predicted a few unstable modes with $k_x\ge 1$, in this restricted domain that does not
allow modes with $k_x<1$ the non-linear coupling together with the truncation of long wavelength modes somehow
resulted in the artificial damping of unstable modes much earlier than expected. 
However, a simple increase in the grid size $L$ by a factor of 2, allowing for 
extra modes to exist ($k_x=0.5$ but also $k_x=1.5,2.5 ...$), immediately changed the picture, allowing the 
perturbations to grow to $\sim B_0$.   
 WH included a large number of modes but with $k_x \ge 1$, thus
the truncation of longer wavelength modes seems to be the reason for the artificial damping of the HI. We note that, as the magnetic field 
geometry deviates from its initial shape, either due to the finite resistivity (the background field is not affected 
by the Hall drift) or because of the growth of the perturbations, the linear analysis is not longer valid.
As the field changes, the $k_x$-window of instability may shift to lower values, and in that case the instability 
will be suppressed unless modes with longer wavelengths are permitted.
Another important feature confirming the validity of our results is that the linear stage always proceeds 
similarly, irrespectively of the size of the initial perturbation (if models are compared with the appropriate 
scaling factor).

\section{Conclusions}

We have proven that there are cases in which the HI dominates the early evolution of the magnetic field and 
completely reorganizes the field structure in a few Hall times. It is not always subsumed by the Hall cascade, as 
argued by other authors. We have found that the finite size of the computational domain, and the use of different 
boundary conditions significantly affect  the outcome.
Since the HI is not entirely a scale-free problem and, when a slab geometry is considered,
the relative sizes of the length and thickness of the slab make a difference, our findings raise new and interesting questions.
In simple terms, a thin slab is  easier subject to the HI than a thicker one but, what happens in a real NS crust ?

The crust of a NS is not an infinite slab, but a finite-size,  curved domain.
To gain insight into realistic scenarios, 
one can mimic the axially symmetric case with a plane parallel model of a longitudinal size $\approx70$ km
and a thickness of 1 km (but with reflecting boundary conditions instead of periodic boundary conditions at the 
poles as used in this work).
This resembles one of our models that has the largest grid size and therefore for which the HI occurs.
By analogy with this toy model, one may be tempted
to claim that the HI is truly relevant and will affect the evolution of the magnetic field in NSs. 
When we compare the structures and evolutions of the models shown in this paper with those of \cite{PonsGeppert2007}, 
both the qualitative shape of the field and the typical evolution are very similar, despite important differences in 
the physical model. 
Perhaps the octupolar structures that survive to late times in the realistic NS crust simulations
of \cite{PonsGeppert2007} originate from the HI.
However, caution is required in accepting this interpretation. Given the sensitivity of the results to details such as the size of the grid 
(which allows only a finite number of unstable modes)
and the existence of non-linear couplings between modes,
it becomes necessary to perform multidimensional simulations in a spherical shell with a 
realistic crust model before we can make any claim about the relevance of the HI in NSs. 
In addition, toy models that have constant density and cubic boxes with 
periodic boundary conditions, such as those studied in WH, cannot be used to argue against the 
relevance of the HI in NSs, which is a global, more complex problem, that also depends on the true dimensions and the 
nature of the physical boundaries. 
As we have proven, the numerical truncation of some modes due to the computational domains being too small can suppress the
instability. An extension of our present code to spherical coordinates and to contain a realistic NS model is under development. 
Very preliminary results seem to indicate that an active Hall stage exists, but only a more detailed
analysis will provide a definitive answer.

\begin{acknowledgements}
We are grateful to M. Rheinhardt and Juan A. Miralles  for providing the vacuum boundary condition and for many enlightening discussions. 
This work was partly supported by CompStar, a Research Networking Programme of the European 
Science Foundation and the Spanish MEC grant AYA 2007-67626-C03-02.
\end{acknowledgements}


\begin{thebibliography}{14}
\expandafter\ifx\csname natexlab\endcsname\relax\def\natexlab#1{#1}\fi

\bibitem[{{Aguilera} {et~al.}(2008{\natexlab{a}}){Aguilera}, {Pons}, \&
  {Miralles}}]{Aguilera2008b}
{Aguilera}, D.~N., {Pons}, J.~A., \& {Miralles}, J.~A. 2008{\natexlab{a}},
  \aap, 486, 255

\bibitem[{{Aguilera} {et~al.}(2008{\natexlab{b}}){Aguilera}, {Pons}, \&
  {Miralles}}]{Aguilera2008a}
{Aguilera}, D.~N., {Pons}, J.~A., \& {Miralles}, J.~A. 2008{\natexlab{b}},
  \apjl, 673, L167

\bibitem[{{Ant{\'o}n} {et~al.}(2006){Ant{\'o}n}, {Zanotti}, {Miralles},
  {Mart{\'{\i}}}, {Ib{\'a}{\~n}ez}, {Font}, \& {Pons}}]{Anton2006}
{Ant{\'o}n}, L., {Zanotti}, O., {Miralles}, J.~A., {et~al.} 2006, \apj, 637,
  296

\bibitem[{{Cumming} {et~al.}(2004){Cumming}, {Arras}, \& {Zweibel}}]{CAZ2004}
{Cumming}, A., {Arras}, P., \& {Zweibel}, E. 2004, \apj, 609, 999

\bibitem[{{Geppert} \& {Rheinhardt}(2002)}]{GR2002}
{Geppert}, U. \& {Rheinhardt}, M. 2002, \aap, 392, 1015

\bibitem[{{Goldreich} \& {Reisenegger}(1992)}]{GR1992}
{Goldreich}, P. \& {Reisenegger}, A. 1992, \apj, 395, 250

\bibitem[{{Hollerbach} \& {R{\"u}diger}(2002)}]{HR2002}
{Hollerbach}, R. \& {R{\"u}diger}, G. 2002, \mnras, 337, 216

\bibitem[{{Hollerbach} \& {R{\"u}diger}(2004)}]{HR2004}
{Hollerbach}, R. \& {R{\"u}diger}, G. 2004, \mnras, 347, 1273

\bibitem[{{Hoyos} {et~al.}(2008){Hoyos}, {Reisenegger}, \&
  {Valdivia}}]{Hoyos2008}
{Hoyos}, J., {Reisenegger}, A., \& {Valdivia}, J.~A. 2008, \aap, 487, 789

\bibitem[{{O'Sullivan} \& {Downes}(2006)}]{OD2006}
{O'Sullivan}, S. \& {Downes}, T.~P. 2006, \mnras, 366, 1329

\bibitem[{{Pons} \& {Geppert}(2007)}]{PonsGeppert2007}
{Pons}, J.~A. \& {Geppert}, U. 2007, \aap, 470, 303

\bibitem[{{Pons} {et~al.}(2007){Pons}, {Link}, {Miralles}, \&
  {Geppert}}]{PonsLink2007}
{Pons}, J.~A., {Link}, B., {Miralles}, J.~A., \& {Geppert}, U. 2007, Physical
  Review Letters, 98, 071101

\bibitem[{{Reisenegger}(2009)}]{Reis2009}
{Reisenegger}, A. 2009, \aap, 499, 557

\bibitem[{{Rheinhardt} \& {Geppert}(2002)}]{RG2002}
{Rheinhardt}, M. \& {Geppert}, U. 2002, Physical Review Letters, 88, 101103

\bibitem[{{T{\'o}th} {et~al.}(2008){T{\'o}th}, {Ma}, \& {Gombosi}}]{Toth2008}
{T{\'o}th}, G., {Ma}, Y., \& {Gombosi}, T.~I. 2008, Journal of Computational
  Physics, 227, 6967

\bibitem[{{Vainshtein} {et~al.}(2000){Vainshtein}, {Chitre}, \&
  {Olinto}}]{Vai2000}
{Vainshtein}, S.~I., {Chitre}, S.~M., \& {Olinto}, A.~V. 2000, \pre, 61, 4422

\bibitem[{{Wareing} \& {Hollerbach}(2009)}]{WH2009}
{Wareing}, C.~J. \& {Hollerbach}, R. 2009, \aap, 508, L39

\end{thebibliography}

\end{document}